# Human Resilience in the AI Era: What Machines Can't Replace

Shaoshan Liu[1], Anina Schwarzenbach[2], Yiyu Shi[3]
[1]Shenzhen Institute of Artificial Intelligence and Robotics for Society (AIRS)
[2]University of Bern
[3]University of Notre Dame

**Abstract:** AI is changing work and decision making faster than many institutions can adapt their operating practices. We argue that this adaptation gap makes human resilience a core capability for the AI era. We define resilience as the capacity to absorb disruption while preserving effective action and human agency around core purposes. The framework operates at three interacting levels. Psychological resilience keeps a person goal-directed under stress. Social resilience makes trusted support and correction available across a group. Organizational resilience turns detected problems into learning and recovery. We connect established resilience and technostress research with direct AI-in-the-loop experiments. General resilience is trainable, while AI-specific causal evidence is still emerging. Direct AI studies show that assistance can raise productivity and spread expertise. Other experiments show improved expressed empathy and more calibrated reliance. We translate these findings into a practical agenda for AI education, workplace design, governance, and evaluation. The central proposal is socio-technical: structural safeguards define the operating boundary, while resilient people and institutions provide adaptive capacity when conditions change.

**Index Terms—**Artificial intelligence, human resilience, human–AI interaction, AI governance, organizational resilience, technostress.

## 1. Can Humanity Keep Up with AI?

Artificial intelligence creates an adaptation-speed problem. Models and autonomous systems can enter workflows faster than organizations can revise how work is organized and governed. That gap matters because AI increasingly operates inside the digital and physical systems through which people already work. In the autonomy economy, increasingly capable machines perform physical and cognitive tasks with less direct supervision [1]. Human outcomes therefore depend on how quickly people and institutions learn to work with these systems while retaining control over consequential decisions.

The disruption reaches beyond productivity. Work provides a livelihood and anchors a sense of competence and purpose within a social role. When AI changes task boundaries or redistributes expertise, workers must learn new forms of contribution while career expectations change around them. Resilience describes the capacity to stay effective through that transition. Technical reskilling supplies new task knowledge; psychological adaptation helps people manage uncertainty and rebuild a coherent sense of direction as the work itself changes.

AI is also changing social interaction. Emotionally responsive systems can generate or recommend empathic language [2]. Human relationships carry a history of reciprocal obligation and accountability. Social resilience therefore concerns the strength of the human network around AI use. The practical objective is to let AI improve communication while preserving relationships that provide trusted help and mutual challenge, with responsibility shared between people.

Judgment provides a third pressure point. AI-assisted decisions can be useful precisely because the system supplies an additional source of advice, yet the same advice can mislead. Gaube et al. asked physicians to interpret chest X-rays while receiving diagnostic advice. Diagnostic accuracy fell when the advice was inaccurate, regardless of whether participants believed the advice came from an AI system or a human expert [3]. Resilience in an AI-mediated decision process therefore means maintaining contestability. Users evaluate advice and can escalate uncertainty into a human review process with authority to override the recommendation and recover from error.

We use human resilience as an organizing lens for these pressures. Human resilience is the multi-level capacity to adapt to disruption and restore effective functioning while preserving core purposes and values. The title uses the language of replacement to make a governance claim: as machine capability expands, responsibility for consequences and for the purposes served by high-impact choices remains attached to people and institutions. The article asks how that responsibility can remain actionable when AI changes the conditions under which people work and decide.

## 2. Human Resilience: Definition and Historical Context

Human resilience operates across three interacting levels. Psychological resilience concerns an individual's ability to remain goal-directed under stress by regulating emotion and revising interpretations or strategies as conditions change [4]. Social resilience concerns access to trusted relationships and shared resources that help a group coordinate during disruption [5], [6]. Organizational resilience concerns the institutional ability to learn from anomalies and restore effective operation while preserving core commitments [7]. The levels reinforce one another. A person adapts more effectively when trusted peers are available, and those peers can act more effectively when the organization gives them permission and procedures to surface problems.

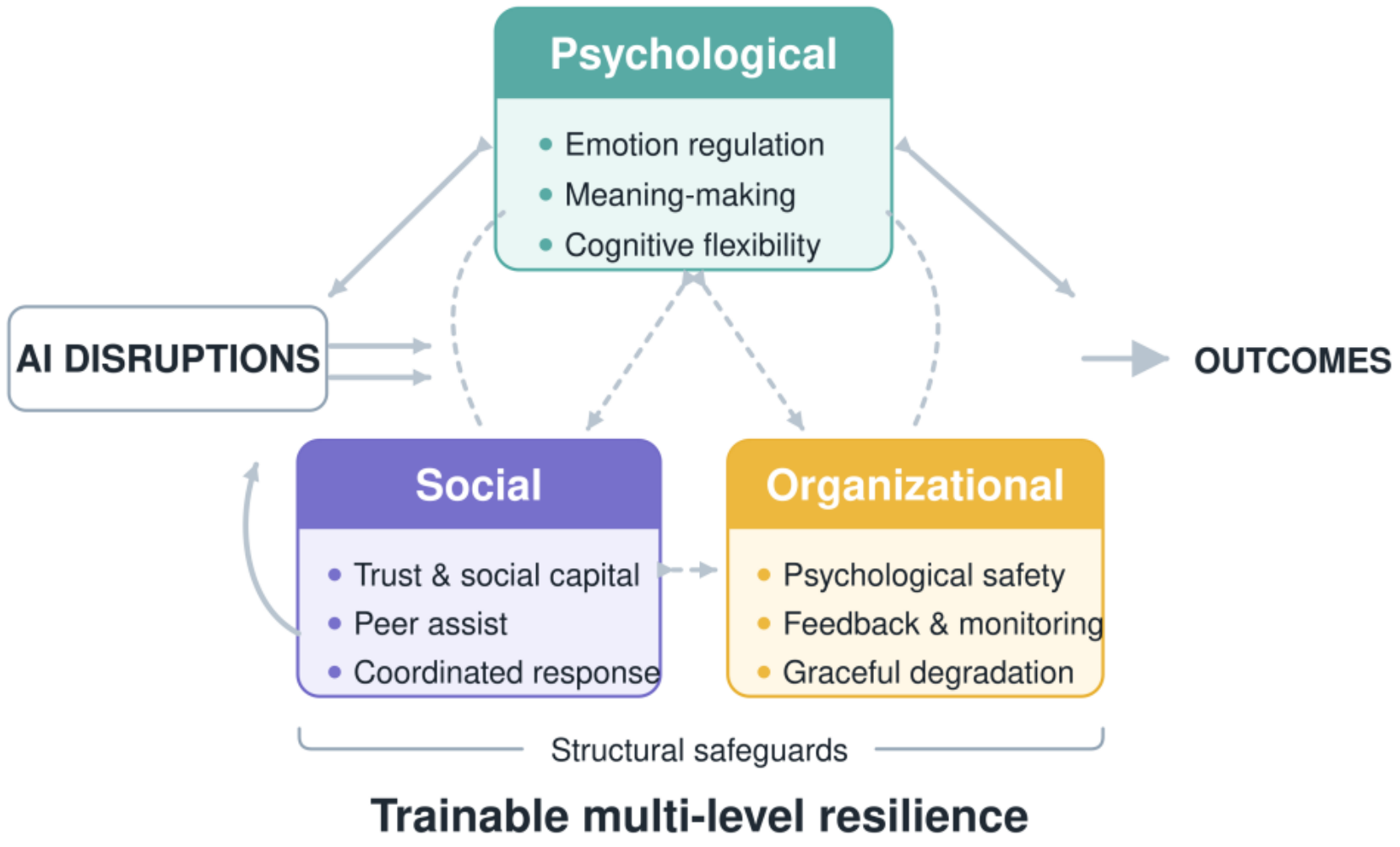

Fig. 1. A multi-level resilience model for AI-mediated disruption. Psychological capacity supports individual adaptation, social capacity enables shared correction, and organizational capacity supports learning and recovery. Structural safeguards define the operating boundary within which these capacities function.

Figure 1 summarizes the mechanism. AI changes the demands placed on a person or workflow. Resilience shapes the response to that change. Psychological capacity helps an individual stay oriented to the task when uncertainty rises. Social capacity makes additional knowledge and correction available through other people. Organizational capacity converts reported problems into learning and recovery. Structural safeguards anchor the model by defining acceptable use and the conditions for oversight and fallback before disruption occurs.

Resilience is a dynamic process shaped by context and available resources. Evidence from earlier disruptions helps establish that general mechanism. During stringent COVID-19 lockdown in Colombia, a 15-day diary study of 109 adults found lower daily distress on days when participants experienced greater social connectedness; daily social connectedness was the significant interpersonal correlate of distress, while associations with the baseline emotion-regulation measures were statistically inconclusive [8]. Disaster research similarly shows how social networks provide information together with practical and emotional support during response and recovery [6]. These studies establish contextual and social mechanisms of adaptation. AI-specific studies are needed to determine how strongly the same mechanisms operate when the disruption comes from automated systems.

## 3. Human Resilience in the AI Era: What the Evidence Shows

The evidence for AI-era resilience comes from two connected bodies of work. Resilience and technostress research provides established mechanisms for adaptation under pressure. AI-in-the-loop studies provide direct observations of behavior when people work with automated advice or generative systems. Most direct studies examine a particular task and measure outcomes such as productivity, reliance, or expressed empathy. We therefore use each study for the mechanism it directly measures. Productivity gains indicate successful tool use; resilience claims require direct measures of psychological, social, or organizational capacity. This distinction keeps the argument proportional to the evidence and defines the next empirical step.

**Psychological resilience.** Digital work can create techno-overload and role stress when systems increase work intensity or complexity. Tarafdar et al. found that technostress was positively associated with role stress and inversely associated with individual productivity in survey data from ICT users across 223 organizations [9]. Later research with 237 institutional sales professionals found that technology self-efficacy and organizational support mechanisms could mitigate performance losses associated with technostress [10]. These results provide a useful baseline for AI-intensive work because rapid model changes and opaque outputs can place similar demands on users who must continually update how they perform a task. The analogy is especially relevant when AI changes the pace or ambiguity of work. In those settings, the user must repeatedly decide whether an unfamiliar output reflects useful novelty, a model limitation, or a change in the task environment. Psychological resilience concerns the capacity to make that adjustment while staying oriented to the task goal.

A 2023 study of 225 remote workers sharpens the mechanism. Techno-stressors were linked to poorer psychological health through burnout, while e-work self-efficacy buffered indirect pathways to depressive mood and anxiety symptoms [11]. Its relevance to AI lies in the role of domain-specific self-efficacy. In an AI workflow, that capacity can be expressed as confidence grounded in skill. The user knows how to inspect an output and revise an approach after a failed interaction. When uncertainty exceeds personal knowledge, the user seeks human expertise.

Direct AI evidence shows how assistance can expand performance. Noy and Zhang ran a preregistered experiment with 453 college-educated professionals completing occupation-specific writing tasks. Participants randomly given access to ChatGPT finished about 40% faster and produced work rated about 18% higher in quality [12]. The study measured productivity rather than resilience, so we use it to show rapid adaptation to a new tool while users retain responsibility for the final task. Participants with ChatGPT were also more likely to report using it in their real jobs after the study [12], suggesting that the experience influenced subsequent behavior. Future longitudinal work can test whether repeated use strengthens independent skill or changes reliance.

Reliance studies reveal the complementary requirement: adaptation must preserve judgment when advice is poor. In the chest-X-ray study by Gaube et al., inaccurate advice reduced diagnostic accuracy across both purported advice sources [3]. Buçinca et al. then tested interface designs that deliberately required more analytical engagement with AI recommendations. In an experiment with 199 participants, cognitive-forcing designs reduced overreliance compared with simpler explainable-AI interfaces, although participants rated the more demanding designs less favorably [13]. The result gives calibrated reliance a concrete role in psychological resilience: a resilient user accepts helpful advice selectively and increases scrutiny when the evidence warrants it.

**Social resilience.** Adaptation is distributed across people because teams collectively carry more context and expertise than any one user. Brynjolfsson, Li, and Raymond studied the staggered deployment of a generative-AI assistant across 5,172 customer-support agents. Access increased issues resolved per hour by 15% on average, with substantially larger gains among less experienced and lower-skilled workers [14]. Treated agents with two months of tenure performed about as well as untreated agents with more than six months, which the authors interpret as evidence that AI assistance can accelerate learning [14]. The highest-skilled workers saw small gains in speed and small declines in quality, showing that the effect depends on user needs. For resilience, the key mechanism is diffusion of useful practice across the team while human mentoring remains available. Employer expectations point in the same direction: in the World Economic Forum's 2025 survey of more than 1,000 employers, "resilience, flexibility and agility" ranked immediately after analytical thinking among core skills [15].

Sharma et al. provide a more explicitly interpersonal case. Their HAILEY system supplied just-in-time writing feedback to 300 peer supporters from the TalkLife community. Human-AI collaboration increased expressed conversational empathy by 19.6% overall and by 38.9% among participants who initially reported difficulty providing support; participants also reported higher self-efficacy [2]. HAILEY supplied feedback to the supporter, who interpreted the situation and chose the response. The result shows how AI can strengthen the quality of human support while judgment, context, and responsibility remain with the person providing it.

Together, these studies suggest a practical principle for social resilience: AI can broaden access to useful expertise while human correction remains easy to reach. Deployment studies can test the social mechanism directly through peer-help behavior and the time required to bring human expertise into a difficult case.

**Organizational resilience.** Institutions create the conditions under which individual judgment and peer correction can influence a system. Psychological safety is one such condition. Edmondson's multimethod field study of 51 manufacturing work teams found that team psychological safety was associated with learning behavior [16]. The AI-era implication is straightforward: organizations need a work environment in which a user can surface an unexpected model behavior or challenge an AI-supported decision with explicit protection for slowing the workflow to raise a safety concern. That reporting channel turns local uncertainty into organizational learning.

The NIST AI Risk Management Framework provides a complementary governance mechanism. Its MANAGE function calls for post-deployment monitoring and explicitly includes appeal and override, incident response, recovery, and change management; it also calls for continual improvement based on measured experience [17]. These practices give organizational resilience an operational form. An AI-enabled workflow can detect a problem, move to a safer operating mode, and incorporate lessons from the incident into subsequent deployment. The framework also links monitoring to decision rights. Detection has limited value when nobody is authorized to pause the system or redirect a case. Resilience therefore depends on both information flow and authority: the organization needs to know that conditions have changed and to empower a person or process to change the operating mode in response. In this article, graceful degradation refers to that controlled movement toward safer operation when confidence falls or an anomaly appears.Table 1 summarizes the key insight at each level and the evidence used to support it.

**TABLE 1. Human Resilience in the AI Era: What the Evidence Shows**

| Resilience level | Evidence summarized | Key insight for AI-era resilience |
|---|---|---|
| **Psychological** | Technostress and self-efficacy research [9]-[11]; AI productivity and reliance studies [3], [12], [13] | AI-era adaptation requires users to gain value from assistance while preserving calibrated judgment when advice is uncertain. Current AI studies mainly measure task performance and reliance, so direct resilience measures remain an empirical priority. |
| **Social** | Generative-AI deployment in customer support [14]; AI-assisted peer support [2] | AI can spread expertise and strengthen human support when human correction remains accessible. Direct team-network measures can test whether this capacity persists as work reorganizes around AI. |
| **Organizational** | Psychological safety in work teams [16]; NIST AI Risk Management Framework [17] | Reporting channels and decision rights can turn detected anomalies into recovery and organizational learning. NIST provides an operational framework; deployment studies can test how these mechanisms affect recovery performance. |

## 4. Can Human Resilience to AI Be Developed?

The practical value of the framework depends on whether resilience can be strengthened. General evidence supports trainability, although the size and durability of effects vary. Robertson et al. reviewed 14 workplace resilience-training studies and found evidence of improvement in personal resilience and employee well-being, with some broader performance benefits. Variation in program design limited conclusions about the most effective format [18]. Vanhove et al. synthesized 42 independent samples from 37 studies and estimated a small overall effect for workplace resilience programs ($d = 0.21$). Their analysis also found a proximal effect of $d = 0.26$ and a distal follow-up effect of $d = 0.07$, showing that maintenance matters [19]. The same meta-analysis found that one-to-one delivery produced the strongest average effects among the formats examined, followed by classroom group delivery, while computer-based and train-the-trainer formats were weaker [19]. These moderator findings should be treated as suggestive because study designs varied, yet they reinforce an important implementation point: resilience is more likely to develop through repeated, context-sensitive practice than through a generic one-time module.

Education provides a second evidence base. A 2025 systematic review identified 38 randomized controlled trials involving 15,730 children and adolescents; 21 trials had data suitable for meta-analysis. School-based interventions produced a small pooled improvement in resilience (SMD = 0.17, 95% CI 0.06-0.29). Heterogeneity was considerable ($I^2 = 81.9\%$), and the authors rated the certainty of evidence as very low [20]. The result supports a specific conclusion: structured intervention can strengthen resilience, but outcomes depend heavily on context and implementation. For AI-era education, schools should define the behavior they want to strengthen, tie training to authentic AI use, and test whether the behavior transfers beyond the exercise.

AI-specific training can begin with judgment under uncertainty. A learner can make an initial assessment before seeing an AI recommendation, record the reason for any change in judgment, and practice recovery after a plausible error. Buçinca et al. support this process design: interfaces that forced more active reasoning reduced overreliance [13]. Repeated practice can vary the reliability of AI advice so that learners experience both strong and misleading recommendations. This gives them direct experience deciding when additional evidence or human review is warranted.

Social resilience requires practice at the team level. Teams can rehearse how a difficult AI-supported case moves from an individual user to a knowledgeable peer and then, when necessary, to a formal escalation channel. A short after-action review can examine where useful expertise came from and how quickly disagreement was resolved. Repeated exercises normalize help-seeking and challenge before a crisis occurs. The exercise can also reveal concentration risk in expertise. If every difficult case eventually reaches the same specialist, the team may appear coordinated while remaining fragile. Social resilience improves when more than one person can provide useful correction and when workers know how to locate that expertise quickly under pressure.

Organizational resilience depends on authority that matches the training. Users need a documented path for human takeover and a clear destination for uncertain cases. High-impact workflows should also rehearse recovery, including periods when the AI tool is unavailable or produces degraded output. The NIST AI RMF already supplies a governance vocabulary for monitoring, appeal, override, incident response, and recovery [17]. Training becomes credible when employees can exercise those mechanisms in the live system.

The evidence supports a clear answer: human resilience can be developed. Workplace and school studies show measurable gains from structured interventions, usually modest and dependent on how programs are implemented [18]-[20]. AI-specific evidence is earlier but already identifies trainable mechanisms: active reasoning can reduce overreliance [13], and organizational practice can rehearse monitoring, override, and recovery using established risk-management processes [17]. Human resilience should therefore be treated as a trainable socio-technical capability. Its development combines individual practice with team support and organizational authority, and future AI-specific trials can determine which combinations produce the most durable effects.

Development should also be evaluated through behavior in the actual workflow. At the individual level, the relevant change is better calibrated judgment under uncertain advice. At the team level, it is faster access to human expertise during difficult cases. At the organizational level, it is a shorter path from a detected anomaly to safe recovery. These outcomes make resilience development observable rather than equating it with completion of a training program.

## 5. From Framework to Practice

Putting the framework into practice means designing AI use so that people can question a recommendation, reach human expertise when needed, and recover when the system fails. The aim is to preserve effective human action as more work is delegated to automated systems. Training, workflow design, and evaluation therefore need to focus on what people actually do when AI is useful, uncertain, or wrong. A resilient deployment is one in which users can take advantage of automation without losing the skills, relationships, or authority needed when conditions change.

In education, resilience is best taught through decisions that contain uncertainty. Students can make an initial judgment, inspect an AI response, and explain what changed their view. Some exercises can deliberately include plausible errors so that detecting the problem becomes part of the task. The useful outcome is calibrated reliance: students learn when AI adds value and when a claim needs more evidence or human review. Task quality can be assessed together with the student's explanation of the final decision and whether the error was recognized. Repeating the exercise across familiar and unfamiliar tasks can also show whether the student has learned a general strategy for evaluating AI output. This makes resilience visible in actual judgment and supports self-report measures with behavioral evidence. It also shifts AI literacy from knowing how to operate a tool toward knowing how to remain responsible for a decision made with that tool.

In workplaces, resilience belongs in workflow design. A high-impact AI-assisted task needs an explicit point where human review enters and a clear route for uncertain cases. Fallback procedures also need to work under real operating pressure, when time is limited and the person best able to help may be occupied elsewhere. Productivity metrics can be combined with near misses, override cases, and recovery time. For less-experienced workers, periodic unaided tasks can reveal whether AI is accelerating learning or masking skill loss. Reviewing overrides can show when users are correcting the model and when the system is creating avoidable friction. Teams can also examine difficult cases after the fact: where did the useful correction come from, how long did it take to reach the right person, and did the same problem recur? The resulting evidence helps managers understand where capability resides and whether the team can continue to function when AI quality falls.

Governance can focus on recovery alongside average model performance. Accuracy and bias tests describe system behavior under specified conditions; resilience becomes visible when those conditions change. An organization can define in advance what happens after repeated disagreement, anomalous output, or a fall in confidence. It can specify where the case goes, who has authority to pause automated action, and which operating mode is used during recovery. A workflow might send repeated user-model disagreement to secondary review, for example, or transfer authority to a human operator after a defined anomaly threshold is crossed. These rules make accountability executable. They also create a record of why intervention occurred, who acted, and how the system returned to acceptable operation. Such records allow organizations to compare incidents over time and determine whether recurring failures are actually being learned from.

Research can test these mechanisms directly. A trial could randomize a cognitive-forcing step or an escalation protocol, introduce controlled AI errors, and record whether users detect them, whether human help arrives, and how long it takes to restore acceptable performance. A second condition can temporarily remove the AI tool to test whether the workflow continues safely. These designs move beyond productivity as the main outcome and ask how the combined human-AI system behaves under stress. Calibrated reliance captures the quality of user judgment. Time-to-help reveals whether useful expertise remains accessible as work becomes more automated. Detection and recovery latency show whether an organization can convert recognition of a problem into action. The measures can be reported separately and then examined together to identify trade-offs, such as a safeguard that improves error detection but slows routine work or a productivity gain that gradually weakens unaided competence.

Time is part of the evaluation problem. Short experiments capture immediate changes in productivity or reliance, while sustained AI use can reshape skill and work routines. Longitudinal evaluation can track unaided competence, transfer to unfamiliar tasks, and access to human expertise after months of AI use. This matters because resilience-training effects can diminish at follow-up [19], and reliance patterns may evolve with experience. If unaided performance improves while assisted performance rises, AI may be transferring useful practice to the worker. If assisted output rises while unaided performance weakens, fallback capacity is eroding. Similar questions arise at the team level when expertise becomes concentrated in a small number of people or when workers stop seeking one another's advice because the AI system has become the default source. These trajectories matter for deployment decisions because the strongest short-term performance can coexist with declining capacity to respond to failure.

Auditability makes the framework operational. Before deployment, organizations can identify the events that trigger review, the people who can change the operating mode, and the evidence that will be retained after an incident. Each intervention then creates a record of what happened, what correction was made, how long recovery took, and whether the same failure recurred. Over time, those records show whether the human-AI system is becoming better at handling disruption rather than merely better at routine operation. Human resilience becomes a practical system property: people remain able to recognize when conditions have changed, mobilize the right correction, and restore safe operation while responsibility for the outcome remains clear.

## 6. Conclusion

AI will continue to change how people work and make decisions. Human resilience provides a framework for keeping those changes governable. Psychological resilience preserves judgment under pressure. Social resilience keeps human expertise and correction available. Organizational resilience turns anomalies into learning and recovery. Structural safeguards set the boundaries within which all three operate.

The evidence is strong enough to support a research and practice agenda with calibrated expectations. Workplace and school studies show that resilience can respond to structured intervention, usually with modest average effects that vary by setting [18]–[20]. Direct AI studies show several mechanisms relevant to that agenda. Generative assistance can raise productivity and spread expertise [12], [14]. AI feedback can strengthen expressed empathy [2], while reliance studies show how inaccurate advice and interface design can shape overreliance [3], [13]. The major empirical opportunity is to measure resilience directly across AI-enabled settings and determine which interventions produce durable effects.

The next stage should treat human resilience as part of AI system evaluation. Trials can test interventions at each level and then observe the combined human-AI workflow under routine operation, controlled error, and recovery. For practitioners, the immediate principle is to pair adaptive capacity with structural safeguards. People need enough skill and authority to recognize when the operating situation has changed; institutions need mechanisms that convert that recognition into safe action. When those elements work together, AI augmentation remains accountable to human purposes.

# Author Biographies

**Shaoshan Liu** is Director of Embodied Artificial Intelligence at the Shenzhen Institute of Artificial Intelligence and Robotics for Society (AIRS). His research interests include embodied AI and technology policy. He received a Ph.D. in computer engineering from the University of California, Irvine, and an M.P.A. from Harvard Kennedy School. He is a Senior Member of IEEE. Contact him at shaoshanliu@cuhk.edu.cn

**Anina Schwarzenbach** is a senior researcher and lecturer at the University of Bern. Her research interests include social cohesion, social networks, resilience, AI and ethics, digital societies, and computational social science. She received a Ph.D. in sociology from the University of Freiburg, an LL.M. in criminology from the University of Bern, and an M.A. in communication science, economics, and criminology from the University of Zurich. Contact her at anina.schwarzenbach@unibe.ch

**Yiyu Shi** is a professor in the Department of Computer Science and Engineering at the University of Notre Dame and director of the Sustainable Computing Laboratory. His research interests focus on hardware intelligence and biomedical applications. He received a B.S. in electronic engineering from Tsinghua University and M.S. and Ph.D. degrees in electrical engineering from the University of California, Los Angeles. He is a Senior Member of IEEE. Contact him at yshi4@nd.edu